\documentclass[11pt,preprint,graphicx]{aastex}

\begin{document} 

\title{Introductory Astronomy as a Measure of Grade Inflation}

\author{James Schombert}
\affil{Department of Physics, University of Oregon, Eugene, OR 97403;
jschombe@uoregon.edu}

\begin{abstract}

We use four years of introductory astronomy scores to analyze the ability
of the current population to perform college level work and measure the
amount of grade inflation across various majors.  Using an objective
grading scale, one that is independent of grading curves, we find that 29\%
of intro astronomy students fail to meet minimal standards for college
level work.  Of the remaining students, 41\% achieve satisfactory work,
30\% achieve mastery of the topics.

Intro astronomy scores correlate with SAT and college GPA.  Sequential
mapping of the objective grade scheme onto GPA finds that college grades
are inflated by 0.2 for natural sciences majors, 0.3 for social sciences,
professional schools and undeclared majors), 0.5 for humanities majors.  It
is unclear from the data whether grade inflation is due to easier grading
curves or depression of course material.  Experiments with student
motivation tools indicates that poor student performance is due to
deficiency in student abilities rather than social factors (such as study
time or decreased interest in academics), i.e., more stringent admission
standards would resolve grade inflation.

\end{abstract}

\section{Introduction and Overview}

One of the most popular science courses taken by non-science majors at
today's universities are introductory astronomy classes.  It is hoped that
non-science students are attracted to the wonder of the cosmos and a wish
to understand the mysteries of the Universe.  However, exit student
evaluations frequently complain of the amount of math and physics in
astronomy courses, suggesting that their initial enthusiasm for the topics
is mediated by the unexpected difficulty of the material.

Despite complaints on exit evaluations, these survey courses continue to be
extremely popular and a significant fraction of non-science majors pass
through intro astronomy to fulfill University science requirements.  Like
all major universities, the University of Oregon also offers a series of
introductory astronomy courses aimed at non-science majors needing to
fulfil general education requirements.  These courses appear to be more
highly attended than other science classes of the same level, although it
is not clear how total enrollment numbers would directly compare to other
science departments since there has never been a term where all the offered
sections were not filled.

The attractive nature of the introductory astronomy courses makes them
ideal laboratories for tracking student cognitive abilities and objective
grading techniques since they address a wide range of students in terms of
ability, class rank and selected major.  They also have a high percentage
of students with undeclared majors, the portion of the student body most at
risk of dropping out of college before completing a degree program.  In
addition, the University of Oregon also has the social science advantage of
admitting a broad range of student aptitude (some faculty would argue the
admission standards are too broad) and a homogeneous student population
that eliminates a majority of ethnic and geographical influences (we are
composed primarily of Caucasian students from the Pacific Northwest).

In this paper, we present a study of student astronomy scores for four
identical classes taught between 2005 and 2009.  The conditions for the
course (exams, quizzes, class size and student distribution) were
unchanging from year to year.  This allows a large sample size to compare
achievement in the course with various cognitive factors such as SAT score
and GPA.  We also apply a unique set of exams to test a students mastery of
the subject material in a objective fashion, independent of the typical
grade curves used to estimate student performance.  This allows us to
compare objectively assigned grades (albeit from a narrowly defined topic,
astronomy) to the wider defined GPA as a measure of grade inflation.  This
technique will also be used to estimate the percentage of students who are
capable of college level work (versus the estimates obtained at admission
standards).  The goal of this project is to present a framework for
objective standards and define preliminary techniques to estimate the
cognitive abilities of the current student population.  

\section{Advantages to ASTR122}

The course selected for our study is ASTR122 (Birth and Death of Stars
course, see http:// abyss.uoregon.edu/$\sim$js/ast122), one of three
introductory astronomy classes taught at UOregon.  There are several
conditions to the ASTR122 course that makes it unique for this type of
study.  For example, economic conditions have eliminated teaching assistant
support for introductory astronomy at UOregon.  The lack of discussion
sections encouraged us to move to an on-line quiz system to supplement the
lecture/exam component.  Thus, the variable of a range in TA quality is
removed and more mathematically intense problem sets were added to the
course material.  On-line material and quizzes means that scores could be
automatically tracked and analysis greatly eased.

ASTR122 presents a more restricted range of physics and math topics than
other sections of introductory astronomy, focused primarily on the stellar
astronomy.  Our other ASTR courses deal with the solar system (thus,
atmospheric chemistry, orbits, gravity, biology, geological processes,
etc.) or cosmology (spacetime, relativity, expanding universe, early
universe processes, particle physics, etc.) and suffer from a broad range of
topics and underlying science.  ASTR122 has the advantage of a more focused
set of physics topics and repeatable mathematical problem sets, leading to
the expectation of more uniformity in student scores.

In addition to the uniformity in course structure and topics, there was an
extra advantage in that (unbeknownst to the students) the exact same exam
and quiz questions were used for all classes over 4 years.  This removes
any chance that, by random question selection, some terms used
particular difficult questions compared to other terms.  This, of course,
brings up the possible cheating; however, there is no indication of that in
the distribution of scores.  Some amount of in-class cheating was detected
on the exams (based on correlated wrong answers on the exams) and these
students withdrew from the course and, thus, the dataset.

\section{Course Structure}

The ASTR122 courses (all taught by the author) were designed to 1) emphasize
mathematical reasoning (with the forum of understanding stellar astronomy)
and 2) present an {\it objective} grading system in order to evaluate the
actual level of college material the current student population can handle.
Mathematical problems were the domain of the on-line quiz system.  Examples
are worked out each lecture (on the blackboard) and the quizzes have
deadlines several days after each lecture allowing for sufficient time for
the students to work out the solutions (and seek assistance).  These were
scored automatically, and the results (plus solutions) posted immediately
after the deadline (providing rapid evaluation for the students).

Knowledge testing was the domain of the multiple choice exams.  Three exams
are taken each term (covering only the previous 1/3 of the course
material).  Each exam is composed of 100 multiple choice questions.  Each
exam is divided into three types of questions.  The first type tests
knowledge of factual information (e.g., what is the color of Mars?).  The
second type of questions addresses the students ability to understand the
underlying principles presented in the course (e.g., why is Mars red?).
The third type of questions examines the ability of the student to process
and connect various ideas in the course (e.g. why is the soil of Mars rich
in heavy elements such as iron?).

Excellence in answering questions of the first type represents satisfactory
grasp of the courses objectives, i.e. a `C' grade.  High performance on
questions of the second type demonstrate good mastery of the course
material, i.e. a `B' grade.  Quality performance on the top tier of
questions would signify superior work, i.e. an `A' grade.  While the design
of the various questions may not, on an individual basis, exactly follow an
objective standard for `A', `B' or `C' work, taken as a whole this method
represents a fairly good model for distinguishing a students score within
what most universities consider a standard grading scheme.  Certainly, this
was the original intent of the ABCDF grading scheme, not to assign a grade
based on class rank or percentage, but to reflect the students actual
understanding of the core material.

With this technique applied to every exam, it is possible to assign an
objective pass/fail line at 45\%. Note, that correctly answering all the
bottom tier questions on a single exam, 33 of them, results in a score of
33\%.  But, the student would then guess the remaining 66 questions for a
total score of 33 + 66/5 = 46\%).  Likewise, the separating point between
`C' and `B' is 60\% and between `B' and `A' is 73\%.  The final grade for
the course is assigned by summing the exam quiz scores and weighting the
exams by 2/3's and the quizzes by 1/3, then normalized to a 100 point
system.  It is these scores that are quoted in the rest of our study as
ASTR122 total score.

\section{Characteristics of the Student Population}

The course was taught every year from 2006 to 2009 by one instructor (the
author).  Enrollment stats are shown in Table 1.  The large number of
withdraws is due to the fact that UOregon allows students to withdraw from
a course up to the 7th week of our 10 week terms (!).  Almost all of the
withdraws would have received failing marks, regardless of their remaining
exam and quiz scores.  In fact, it is surprising that any students fail
this course as by the 7th week they have completed two exams and 2/3's of
the quizzes. This should provide the student with a remarkably accurate
estimate of their final grade (95\% of students failing by the 7th week
fail the course).  Interviews with failing students indicates their
decision to remain in the course is a mixture of 1) a poor understanding of
the arithmetic behind the grading scheme, 2) a overly optimistic opinion of
their ability to overcome a failing score with the last exam and quizzes
and 3) complete disregard of their scores and the impact on their grade for
the course.

Unsurprisingly, the classes were primarily composed of lower classmen
(freshmen and sophomores, 69\%) with equal numbers of juniors and seniors
(see Table 1) as is typical for survey courses.  The majors for the
population, divided crudely into five categories, are 1) 54 (7\%) natural
sciences, 2) 128 (16\%) social sciences, 3) 255 (33\%) professional
schools, 4) 135 (17\%) humanities and 5) 211 (27\%) undeclared.  This
distribution is driven, primarily, by the university general education
requirements for non-science majors.  The high number of undeclared majors
reflects the predominant lower classmen component to the classes.

\begin{deluxetable}{lrrrrr}
\tablecolumns{6}
\small
\tablewidth{0pt}
\tablecaption{Enrollment ASTR122 by term}
\tablehead{
\colhead{} &
\colhead{Fall 07} &
\colhead{Fall 08} &
\colhead{Spring 08} &
\colhead{Fall 09} &
\colhead{Total} \\
}
\startdata

Freshmen  &  79 &  76 &  57 &  89 & 301 (38\%) \\
Sophomore &  56 &  51 &  73 &  63 & 243 (31\%) \\
Juniors   &  23 &  34 &  34 &  33 & 124 (16\%) \\
Seniors   &  22 &  24 &  45\tablenotemark{a} &  24 & 115 (15\%) \\
Total     & 180\tablenotemark{b} & 185\tablenotemark{b} & 209 & 209 & 783 \\

\enddata
\tablenotetext{a}{\small A surge of seniors in the spring term reflects a
number of graduating students desperate to complete science requirements
before spring graduation.}
\tablenotetext{b}{\small A decrease in total enrollment was due to a 
smaller classroom}
\end{deluxetable}

\begin{deluxetable}{llrrrrr}
\tablecolumns{6}
\small
\tablewidth{0pt}
\tablecaption{Score Distribution by term}
\tablehead{
\colhead{} &
\colhead{} &
\colhead{Fall 07} &
\colhead{Fall 08} &
\colhead{Spring 08} &
\colhead{Fall 09} &
\colhead{Total} \\
}
\startdata

Withdraws & F (failure) & 19 & 20 & 35 & 30 & 113 (14\%) \\
$<$45 & F (failure) & 23 & 39 & 30 & 24 & 116 (15\%) \\
45 to 60 & C (satisfactory)  & 79 & 75 & 83 & 89 & 326 (41\%) \\
60 to 73 & B (good) & 41 & 38 & 43 & 56 & 178 (22\%) \\
$>$ 73 & A (superior) & 16 & 13 & 19 & 13 &  61 (8\%) \\

\enddata
\end{deluxetable}

SAT$_M$ and SAT$_R$ scores for the student population are shown in Figure
1 as normalized histograms (592 students).  ACT scores were converted to
SAT equivalents using standard ACT to SAT concordance tables.  The peaks
and means were identical from course to course.  The SAT$_M$ peak is
slightly higher than SAT$_R$, which is unusual in that, for the general
student population, this trend is reversed.  It seems to indicate that
students who elect to take introductory astronomy are slightly more math
proficient that the typical university student.

\begin{figure}
\centering
\includegraphics[scale=0.6,angle=-90]{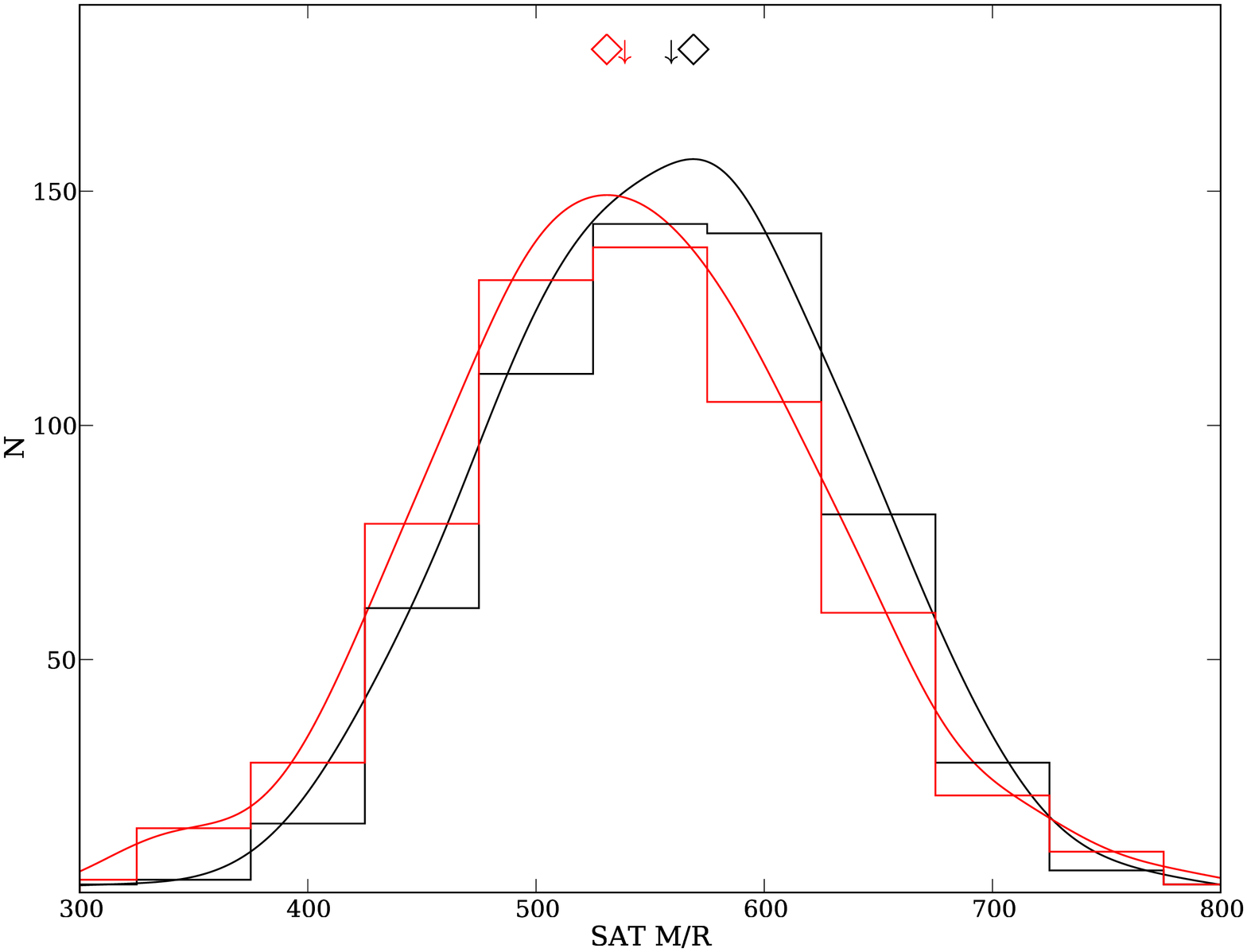}
\caption{SAT Math and Reading for the ASTR122 students.  The math scores are
slightly higher than the general student population (perhaps a natural
selection away from physical sciences by students with weak math scores).  }
\end{figure}

\section{Analysis}

The normalized histograms of the ASTR122 total scores for each class are
shown in Figure 2.  The means were 56.5, 54.1, 56.7 and 56.3 with standard
deviations of 12.  The peak scores were also nearly identical.  The
identical mean scores are not too surprising given the uniformity of the
course structure and the identical nature to the quizzes and exams.
However, total University enrollment numbers increased by over 30\% during
the same time period, so there was some expectation that mean scores would
decrease (as the timescale for human evolution of intelligence was longer
than the increase in enrollment numbers).  We note that given the
definition of our grading standards, these mean ASTR122 scores correspond
to approximately a C+ (versus the common B- used in grading curves).

\begin{figure}
\centering
\includegraphics[scale=0.6,angle=-90]{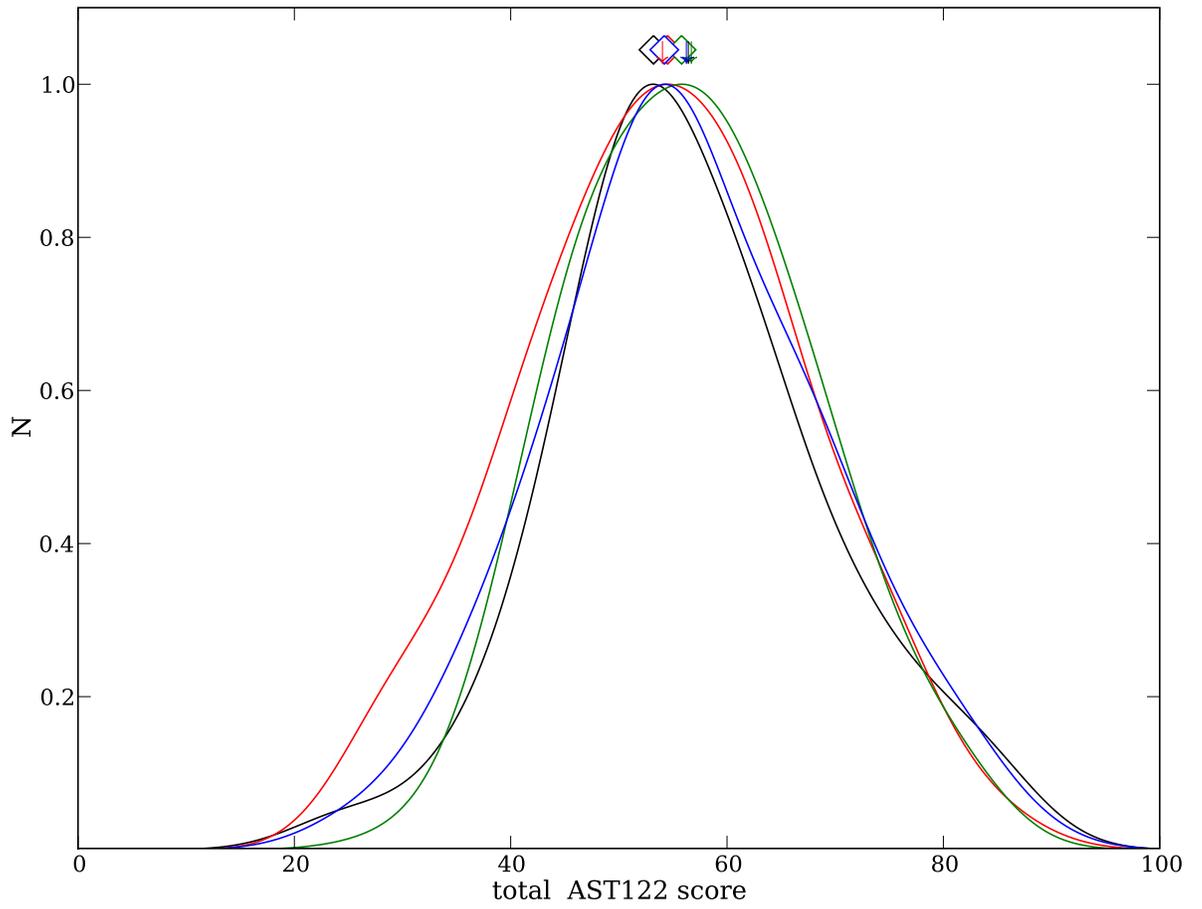}
\caption{The normalized histograms for total course score in ASTR122 for the
four terms listed in Table 1.  Peaks, means and variance are identical to
within the errors.}
\end{figure}

Given the attempt to design the course to objectively assign
`satisfactory', `good' and `superior' level work, we can convert these
total scores into those categories.  To this end, we assign failure, as
outlined in the last section, to scores less than 45\%.  Satisfactory work
(i.e. `C' level) falls in the range of 45\% to 60\% (i.e. 1/2 the distance
between the failure line and the 2/3'rds line for `B' level questions,
adjusted for guessing, 66.6 - 33.3/5).  Good work falls between 60\% and
73\% (i.e. `B' level).   And superior work lies above 73\% (i.e.  `A'
level, 66.6 + 33.3/5).

Table 2 displays the score breakdown for all 794 students.  Withdraws were
treated as failures.  While there are cases where a student with a high
score was forced to withdraw for non-academic reasons, this was rare.  A
vast majority of the withdraws had failing scores up to the point of
withdraw.  Combined withdraws and failures means that 29\% of the students
did not reach the minimal requirements for satisfactory understanding of
the material.  About 41\% of the students achieved a level of
`satisfactory' performance based on exams and problem sets.  Only 22\%
achieved `good' or `B' level work and an additional 8\% reached a level of
`superior' or `A' level work.

If these percentages reflect overall college performance, we are faced with
a scenario where almost 1/3 of admitted freshmen are ill prepared for
college level work (even at an introductory level) or, if they have college
skills, fail to apply them.  Those numbers match freshmen retention and
graduation rates for many universities (the University of Oregon being at
the low end of AAU institutions where retention rates are 80\%); however,
these numbers directly impact on admission and retention strategies as our
data indicates that financial or social reasons are not the primary reason
for student's withdrawing from college.

In addition, only 1/3 of the student population seem capable of `mastery'
level work (above the `B' level), even for an introductory course.  The
remaining population does adequate work, which will probably continue into
their upper division courses (see \S6).  Whether the reasons for this are
due to intelligence, work ethic or simple motivation to perform can not be
fully addressed by this study, but, again, all indicators are that many
undergraduates do not obtain a full education under the current higher
education system.

Mean total scores do not vary between students majoring in humanities, social
science, professional school or undeclared (means of 55$\pm$12).  However,
natural science students have a mean score of 61$\pm$13, significantly
higher than other majors.  This is probably due to an higher familiarity
with math and science terminology; however, similar scores across other
majors will be important in the comparison between scores and GPA as a
measure of grade inflation (see \S7)

\begin{figure}
\centering
\includegraphics[scale=0.8]{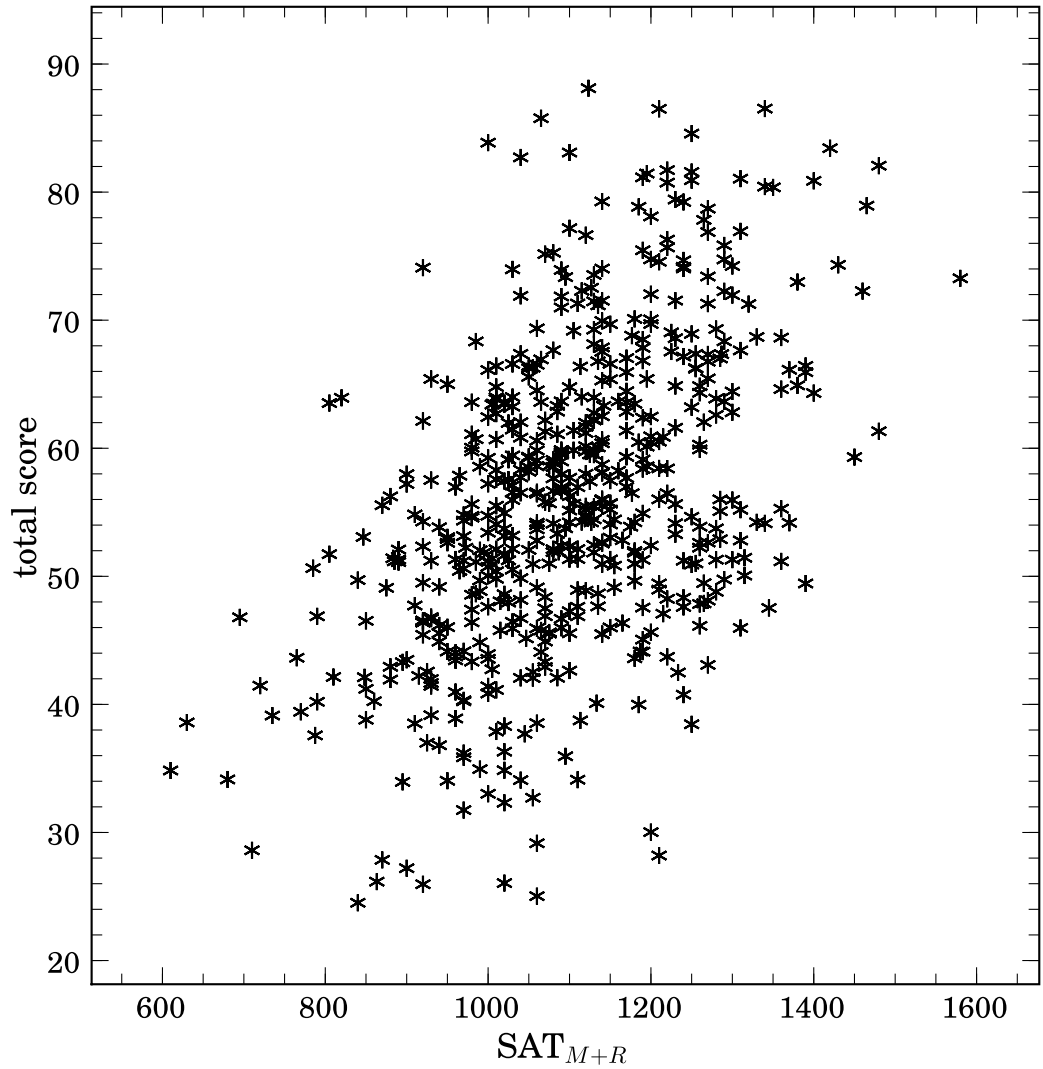}
\caption{Combined SAT score versus ASTR122 score.  The correlation between
SAT score and ASTR122 score is obvious.}
\end{figure}

\section{SAT Scores, GPA and ASTR122}

We can investigate the effect of IQ or `g' component values on class scores
by comparing SAT scores to ASTR122 total scores.  A plot of composite SAT
score (math plus reading) versus total course score is shown in Figure 3.
The correlation is evident (R=0.49) and also visible in SAT$_M$ versus
course score (less obvious in SAT$_R$ versus course score which is
unsurprising due to the mathematical nature of the course).  Due to the
large number of data points, a better method to display the dataset is with
a density distribution, where the data points are binned on a 2D grid and
number of points determines the intensity of each grid pixel.  The density
distribution for SAT scores is shown in Figure 4, along with a moving
average (red symbols) and the peak values in SAT bins of 100 (green line).
While there is a range of scores in each SAT bin, the trend for increasing
ASTR122 score with increasing SAT score is obvious.  The upper left portion
of this diagram is nearly empty, indicating that even extremely
conscientious, hard-working students can not overcome the factors that
produced their low SAT scores.

This is dramatically different from our comparison of SAT scores and upper
division GPA (by major) in our study of over 4,000 students across 17
majors (Hsu \& Schombert 2010).  In that study, there were a significant
fraction of students who obtain high GPA's (greater than 3.5) out of
proportion to their SAT scores.  Presumably, they achieve this academic
success through diligence and hard work; however, these same factors should
apply to introductory astronomy students.  And yet, the ASTR122 scores are
much more strongly correlated to an IQ measurement than any other college
performance indicator.

There are several possible reasons for the difference in SAT to GPA/score
correlations.  For example, the very nature of science and mathematical
knowledge may form a gap between a students cognitive ability and the
threshold needed to master a science topic.  No amount of hard work can
overcome deficiencies in science background that have their origin from
primary and secondary education.  Another possibility is that students in
introductory courses are simply not motivated to work as hard as they do
for courses in their major.  Upper division courses also naturally have a
population of students who have completed their first couple years of
study.  Introductory survey courses have a large fraction of the students
who will dropout by their junior year.  And this population comprises a
majority of the lower tail of the grade distribution where the problem
lies.

\begin{figure}
\centering
\includegraphics[scale=0.8]{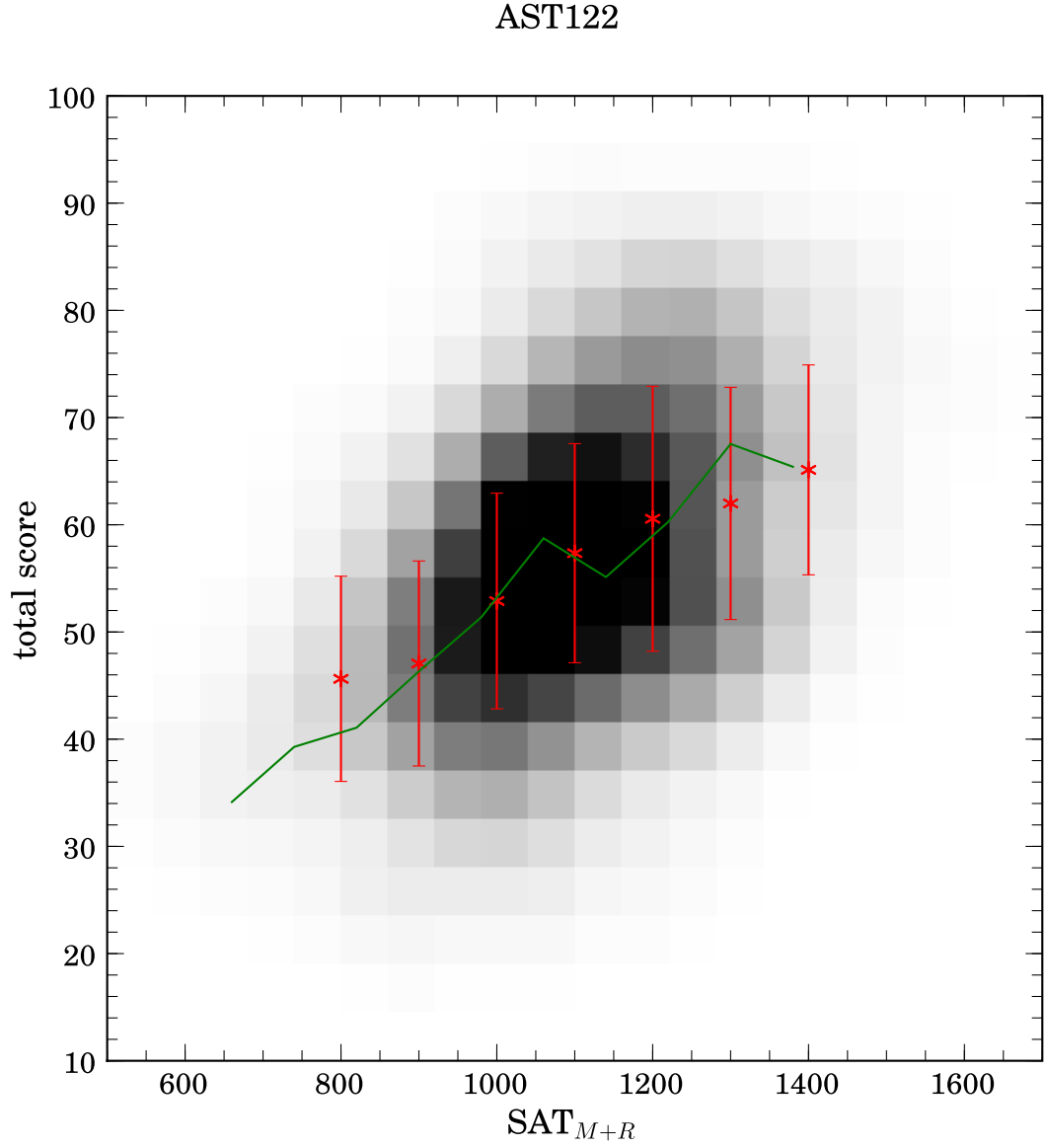}
\caption{Combined SAT score versus ASTR122 score, density distribution.  Shaded
pixels are the data density, red points are the moving average and the
green line is the peak (in bins of 75).  All three demonstrate the strong
correlation between SAT score and ASTR122 course score.}
\end{figure}

One could argue that the results from this study only apply to a particular
scenario, non-science majors performance in an astronomy course.  However,
the performance, as measured by course score, is also highly correlated
with overall college GPA.  Figure 5 displays this data, again as a density
distribution due to the large number of data points.  The Pearson
correlation coefficient is 0.65, which is as high than any know `g'
component correlation to college scores.  It seems, by some unknown
factors, that ASTR122 is a high predictor of total college performance,
stronger than SAT or ACT scores (Hsu \& Schombert 2010). 

\begin{figure}
\centering
\includegraphics[scale=0.8]{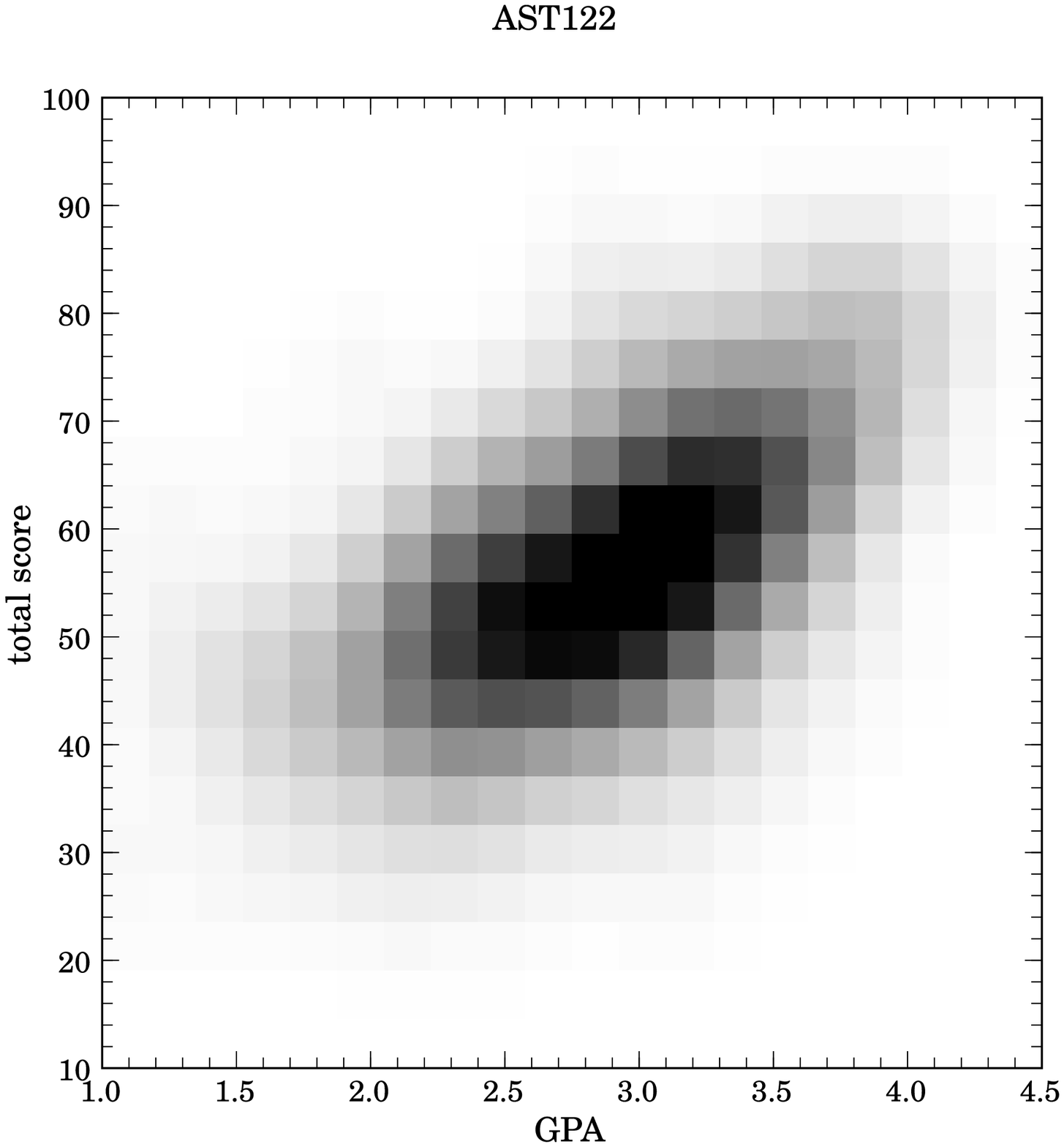}
\caption{Overall GPA versus ASTR122 course score.  The correlation between
ASTR122 and overall GPA is stronger than any predictor of college
performance.}
\end{figure}

Figure 5, and the correlation with SAT score, raises an extremely
disturbing possibility; that survey courses of this type are not evaluating
a students ability to learn the material (in this case, stellar astronomy)
but rather serving as massive IQ tests (fine tuned for college skills).
While one expects smarter students to perform at a higher level, there is
no evidence in these scores that students perform above their expected
levels (i.e. they become educated by the course material) as is seen in
upper division GPA by the time they graduate.

\section{Grade Inflation versus Course Content Depression}

Clearly the distribution of objective grades outlined in the previous
section would be unacceptable to the student population (probably
university administrators and parents as well).  It is these same social
forces that drive grade inflation (outlined in numerous studies, see
http:// gradeinflation.com).  We will not elaborate on these forces here.
Instead, we can map our expected grades onto overall GPA as a measure of
the effects of grade inflation to the general student population.  Note
that this mapping assumes that the objective grading method, outlined in
\S2, does accurately measure a standard that would be agreed upon by most
university educators as consistent with the current interpretation of 'A'
through 'F'.  This is the premise for this entire study and is only
supported by the tight correlations between ASTR122 total scores and SAT
score/GPA.

Using Figure 5, and a least square fit to the moving average, we find that
the failure level (45\%) maps into an overall GPA of 2.4.  Satisfactory
work occurs between 2.4 and 3.1, good work occurs between 3.1 and 3.7 and
excellent work occurs above 3.7.  This would indicate an average grade
inflation of about 0.4 at the low end of the grade scale, decreasing to 0.3
at the high end.  This is consistent with the estimated mean inflation rate
of 0.4 since the 1960's for public universities.

\begin{deluxetable}{lccc}
\tablecolumns{6}
\small
\tablewidth{0pt}
\tablecaption{Grade Inflation by Area of Study}
\tablehead{
\colhead{Major} &
\colhead{'C'} &
\colhead{'B'} &
\colhead{'A'} \\
}
\startdata

Natural Sciences     & 2.2 & 3.1 & 3.5 \\
Social Sciences      & 2.3 & 3.2 & 3.7 \\
Humanities           & 2.6 & 3.5 & 3.9 \\
Professional Schools & 2.3 & 3.2 & 3.7 \\
Undeclared           & 2.3 & 3.2 & 3.7 \\

\enddata
\end{deluxetable}

Grade inflation is also slightly correlated with major.  Again, using the
divisions outlined in \S2 (natural sciences, social sciences, humanities,
professional schools, undeclared), we found the natural sciences to only
have a grade inflation rate of 0.2 at the low end, and none at the higher
grades.  The social sciences, professional schools and undeclared majors
display grade inflation from 0.3 at low grades to 0.2 at the high end.  The
humanities suffered the most extreme grade inflation of 0.6 at the low end
to 0.4 at the upper grades (see Table 3).  While one might argue that the
extreme grade inflation in humanities is not due to changes in GPA, but low
ASTR122 scores, remember the mean ASTR122 scores for all the majors
(excluding the natural science majors) was identical.  A similar trend was
also detected in SAT scores versus GPA (Hsu \& Schombert 2010), where the
humanities displayed higher GPA's per SAT bin compared to other majors.

While one could interpret this shift in objective scores to grade inflation
of overall GPA, another possibility is that this change is due to course
content depression.  Every effort was made to keep the ASTR122 material at
a college level, meaning that all the work assumes high school levels of
math and science.  Thus, the mismatch from ASTR122 score to overall GPA may
reflect that, on average, most college courses have lowered the rigor of
their subject material.  Higher grades, then, are due to better student
performance with lighter material, not due to lower grading standards.  The
reality of the day-to-day practicalities of teaching suggests that both
effects come into play.  A majority of instructors use grade curves, with
little connection to any objective standard.  We simply assume that
admission standards produce a college-ready population of students and
curve the grades to match that standard.  In addition, instructors do wish
to educate.  Therefore, if the course content is not being absorbed, then
lowering the content is a natural (and appropriate) response.

\section{Conclusions}

\noindent We can summarize our findings as the following:

\begin{itemize}

\item{} A series of unique conditions allowed four terms of introductory
astronomy at the University of Oregon to be used to study college
performance for the current student population.  Those conditions are 1)
highly repeatability course structure and material, 2) exams which are
specifically designed with an objective grading scheme and 3) a homogeneous
collection of students across a spectrum of majors.

\item{} Total scores for ASTR122 were uniform from term to term with
identical means and standard deviations.  Natural science majors scores
slightly higher than other majors.  But, means and distributions were
identical for social science, professional school, humanities and
undeclared majors.

\item{} Mapping ASTR122 total scores onto a 'F' (failure), 'C'
satisfactory, 'B' (good) and 'A' (superior) grading scheme, we found that
29\% of college students did not achieve a passing score in these classes.
About 41\% received satisfactory marks.  Only 22\% and 8\% obtained good
and superior marks.

\item{} Total ASTR122 scores correlates well with intelligence factors such
as SAT score.  Total ASTR122 score also correlates extremely well with
total college GPA.

\item{} Mapping objective ASTR122 scores onto overall GPA provides a
measure of global grade inflation.  The resulting values by major are 0.2
for natural sciences majors, 0.3 for social sciences, professional schools
and undeclared majors), 0.5 for humanities majors.

\end{itemize}

In general, the performance of non-science majors in an introductory course
is less than optimal.  This could be due to a number of factors, but all of
them reduce to two possible scenarios: 1) the University admits student
lacking the skills to perform basic college level tasks or, 2) students are
able to perform college level work, but lack the motivation or time to do
so.  In other words, the students either have the skills or they do not.
If they have the skills, then the problem becomes mostly one for the
students to address, i.e., applying their skills to college level material.
Grade inflation weakens the 'stick' for that positive reinforcement cycle.
If they lack the skills, then our mission becomes more difficult, basically
adapting our teaching to skill building rather than knowledge building
(i.e., doing what the secondary schools failed to do).

\begin{figure}
\centering
\includegraphics[scale=0.6,angle=-90]{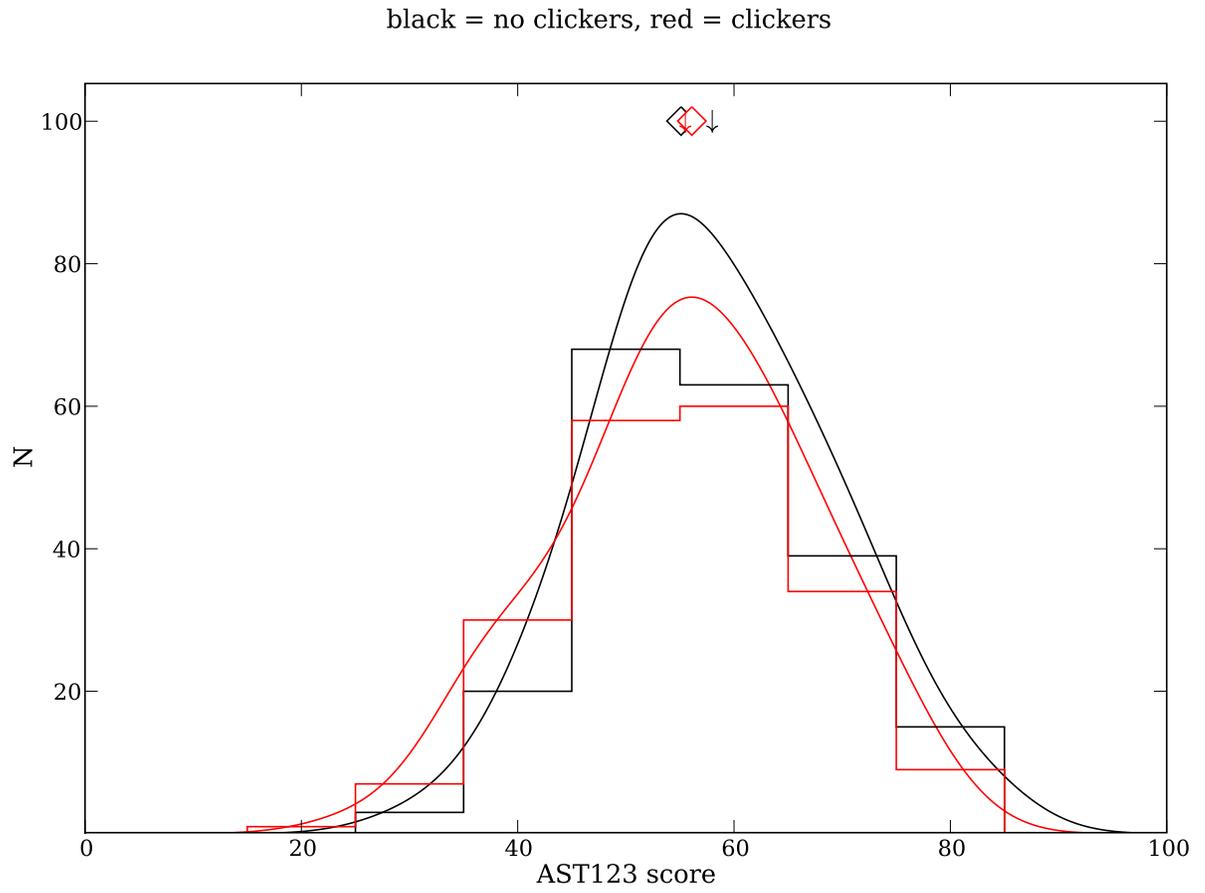}
\caption{Total course scores for an introductory astronomy class
(cosmology) with and without the use of clickers. The was no significant
difference in exam and on-line quiz scores.}
\end{figure}

As a quick test of the hypothesis on low student motivation, we introduced
the use of clickers in the 2009 academic year.  Clickers are the
pedagogical toy-du-jour.  Their usage is thought to increase student
involvement in the course material and lecture attendance.  Both of which
should significantly improve student performance as measured by exam and
quiz scores.  To this end, we experimented on another introductory
astronomy course (cosmology) over two terms, one with and one without the
use of clicker technology.

Between these two courses, which taught identical material, class
attendance increased from a mean of 40\% per lecture to 85\% for the
clicker course.  Student evaluations increased 30\% for the clicker course.
We consider these to be the two key elements in student involvement;
positive student evaluations and high class attendance.  However, Figure 6
displays the total scores for the two classes.  Regardless of the
significant improvement in the classroom environment, the exam and on-line
quiz scores were identical for both terms (in fact, there was a slight
increase in students with failing scores for the clicker class).  This
would indicate that the problem with lower student performance is
ultimately tied to their ability and skill level, not motivation or social
factors.

\acknowledgements

This type of study would not be possible without the development of
network tools which can access and parse data webpages.  Whether the data
is in HTML or SQL format, the extraction of thousands of student records
requires fairly intelligent, non-interactive scripts.  These very tools
were first invented for a knowledge based project funded by NASA's AISR
program.  Their flexibility allows for a wide range of new avenues of
inquiry across the fields of natural and social sciences, the kind of
e-science demonstrated by ongoing efforts for the National Virtual
Observatory.

\end{document}